# A Metric Between Probability Distributions on Finite Sets of Different Cardinalities and Applications to Order Reduction

M. Vidyasagar, Fellow IEEE*

June 6, 2018

**Abstract**

With increasing use of digital control it is natural to view control inputs and outputs as stochastic processes assuming values over finite alphabets rather than in a Euclidean space. As control over networks becomes increasingly common, data compression by reducing the size of the input and output alphabets without losing the fidelity of representation becomes relevant. This requires us to define a notion of distance between two stochastic processes assuming values in distinct sets, possibly of different cardinalities. If the two processes are i.i.d., then the problem becomes one of defining a metric between two probability distributions over distinct finite sets of possibly different cardinalities. This is the problem addressed in the present paper. A metric is defined in terms of a joint distribution on the product of the two sets, which has the two given distributions as its marginals, and has minimum entropy. Computing the metric exactly turns out to be NP-hard. Therefore an efficient greedy algorithm is presented for finding an upper bound on the distance. We then study the problem of optimal order reduction in the metric defined here. This problem also turns out to be NP-hard, so again a greedy algorithm is constructed for finding a suboptimal reduced order approximation. Taken together, all the results presented here permit the approximation of an i.i.d. process over a set of large cardinality by another i.i.d. process over a set of smaller cardinality. In future work, attempts will be made to extend this work to Markov processes over finite sets.

# 1 Introduction

## 1.1 Motivation

As digital control increasingly replaces analog control, in many situations it is more logical to view various signals as discrete-valued, assuming values over a finite alphabet, rather than signals assuming values in some Euclidean space. Suppose we view a control system as an input-output map where the input signal is a sequence $\{u_t\}$ assuming values in some finite set $U$, while the output signal is a sequence $\{y_t\}$ assuming values in another finite set $Y$. In this setting, the problem of order reduction is quite different in nature from the traditional order reduction problem, where the emphasis is on reducing the dimension of the (Euclidean) state space. For the latter problems, well-established methods such as balanced truncation, optimal Hankel norm reduction etc. are appropriate. However, in the case of discrete-valued signals, a different paradigm is required.

*Cecil & Ida Green Chair, Erik Jonsson School of Engineering & Computer Science, The University of Texas at Dallas, 800 W. Campbell Road, Richardson, TX 75080; email: M.Vidyasagar@utdallas.edu. This research was supported by National Science Foundation Award #1001643.

If the system has some element of randomness in it, we should view $\{(u_t, y_t)\}$ as a stochastic process assuming values in the set $U \times Y$.[1] For the purposes of controller design, it would be worthwhile to know whether the finely quantized inputs and outputs can be replaced by a coarser quantization without losing too much accuracy in the representation. Such considerations become particularly germane in the problem of control over networks, whereby the plant and controller may be connected only through a noisy channel. This type of order reduction would require approximating the original stochastic process by another one assuming values in a set of smaller cardinality $U' \times Y'$. The approximation can be quantified by defining a metric distance between two stochastic processes assuming values in distinct sets (of different cardinalities). So far as the author is aware, no such metric is available in the literature. The closest the author has been able to find is a paper by Ornstein [15] in the inaugural issue of the *Annals of Probability*, in which he defines a metric distance between two stochastic processes assuming values in a *common finite set*. Thus the author was motivated to study the problem of defining a metric distance between two stochastic processes assuming values in two distinct sets, as a long-term project.

Naturally, this general problem is very difficult. As a first step, in this paper we tackle the problem of defining a metric between two i.i.d. processes $\{X_t\}$ and $\{Y_t\}$ assuming values in two distinct sets $\mathbb{A}$ and $\mathbb{B}$. Since an i.i.d. process is completely characterized by its one-dimensional marginal, the problem now becomes the following: *Given two probability distributions $\boldsymbol{\phi}$ on $\mathbb{A}$ and $\boldsymbol{\psi}$ on $\mathbb{B}$, can we define a distance $d(\boldsymbol{\phi}, \boldsymbol{\psi})$ that satisfies the usual requirements of symmetry and the triangle inequality?* Let us suppose we succeed in this endeavor. The next logical question would be the optimal order reduction problem, defined as follows: Suppose $\boldsymbol{\phi}$ is a probability distribution on a set $\mathbb{A}$ of cardinality $n$, and suppose an integer $m < n$ is specified. What is the distribution $\boldsymbol{\psi}$ on a set of cardinality $m$ that is closest to $\boldsymbol{\phi}$ in the metric $d$? This question would be very natural in the context of reduced-order modeling of quantized noise, where the assumption that the noise is an i.i.d. process is not so unrealistic. The optimal order reduction problem therefore consists of optimally approximating a quantized noise with a large number of values by another one with far fewer possible values.

---

[1]By this we mean that at each instant of time $t$, $(u_t, y_t)$ belongs to the set $U \times Y$.

Since our motivations include data compression, transmission over noisy channels, etc., it is natural to draw upon well-developed techniques from information theory to achieve our objectives. The use of information-theoretic methods in the controls community has a long history, going back at least to [16] if not much earlier. In this paper, we first define a metric distance between two distributions on distinct finite sets by maximizing their mutual information. Second, we show that, given a distribution $\boldsymbol{\phi}$, all optimal reduced-order approximations of $\boldsymbol{\phi}$ must be aggregations of $\boldsymbol{\phi}$, that is, obtained by adding together various components of $\boldsymbol{\phi}$. It turns out that actually computing the metric distance between two probability distributions, and computing the optimal reduced order approximation, are both NP-hard problems, because they can both be reduced to nonstandard bin-packing problems. Therefore we develop efficient greedy algorithms for both problems. Specifically, we can compute an upper bound on the distance in $O((n+m^2)\log m)$ operations where $n$ and $m$ are the cardinalities of the two sets with $n \geq m$, and we can compute a suboptimal reduced-order approximation in $O(n \log m)$ operations. Note that both complexities are linear in the size of the larger distribution.

As stated above, we view this paper as just the first step in a program. It may be possible to extend the approach proposed here to define a metric distance between two Markov processes assuming values in two distinct sets, and to approximate optimally a Markov process with a large state space by another with a much smaller state space. Those are all questions for future research.

## 1.2 A General Observation

If $\boldsymbol{\phi}, \boldsymbol{\psi}$ are both probability distributions on a common (finite) set $\mathbb{A}$, then a popular and meaningful metric between them is the **total variation metric** defined by[2]

$$\rho(\boldsymbol{\phi}, \boldsymbol{\psi}) = \max_{S \subseteq \mathbb{A}} |\boldsymbol{\phi}(S) - \boldsymbol{\psi}(S)| = \frac{1}{2} \sum_{i \in \mathbb{A}} |\phi_i - \psi_i|.$$

However, if $\boldsymbol{\phi}, \boldsymbol{\psi}$ are probability distributions on *two different sets* $\mathbb{A}, \mathbb{B}$ of different cardinalities, there is no obvious way to define a notion of distance between them. To illustrate, suppose $\mathbb{A} = \{T, F\}$ to suggest 'True' and 'False', while $\mathbb{B} = \{He, Ta\}$ to suggest 'Head' and 'Tails'. Suppose $\boldsymbol{\phi} = [0.3\ \ 0.7]$ and $\boldsymbol{\psi} = [0.8\ \ 0.2]$ are probability distributions on $\mathbb{A}$. Then they are clearly very far apart. Since the set $\mathbb{A}$ consists only of abstract symbols, we can permute the order of the symbols and write $\mathbb{A} = \{F, T\}$. In this case the probability distributions $\boldsymbol{\phi}, \boldsymbol{\psi}$ would also get permuted to $\boldsymbol{\phi} = [0.7 \ 0.3]$ and $\boldsymbol{\psi} = [0.2 \ 0.8]$ respectively, and they are still far apart. In other words, the total variation metric has the natural property that it is permutation-invariant, provided *the same* permutation is applied to the components of both distributions. Now suppose $\boldsymbol{\phi}$ is a distribution on $\mathbb{A}$ while $\boldsymbol{\psi}$ is a distribution on $\mathbb{B}$. Are they close or are they far apart? If we were to make the association $T \leftrightarrow He, F \leftrightarrow Ta$, then they are very far apart, whereas if we were to make the association $T \leftrightarrow Ta, F \leftrightarrow He$, then they are very close. Note that, since the two sets $\mathbb{A}$ and $\mathbb{B}$ are distinct, even if they have the same cardinality, neither association is more natural than the other one. As a result, whatever metric we define between probability distributions on distinct sets (even of the same cardinality), it must be permutation-invariant even if *two different* permutations are applied to the components of the two distributions. In particular, if we permute the components of *only one* of the two distributions, the distance must remain invariant. Note that this is not an artifact of a particular definition. Rather, it is an essential requirement that any definition must satisfy. One consequence is that *any* definition will not reduce to the total variation metric even if $|\mathbb{A}| = |\mathbb{B}|$.

---

[2]The Kullback-Leibler divergence is neither symmetric nor does it satisfy a triangle inequality. Hence it is not a metric, though people often refer to it as a metric or a distance.

## 1.3 Contributions of the Paper

In this paper, we define a distance $d(\boldsymbol{\phi}, \boldsymbol{\psi})$ between two probability distributions $\boldsymbol{\phi}, \boldsymbol{\psi}$ on distinct sets $\mathbb{A}, \mathbb{B}$ by choosing the joint distribution $\boldsymbol{\theta}$ on $\mathbb{A}, \mathbb{B}$ respectively, so that $\boldsymbol{\theta}$ has marginal distributions $\boldsymbol{\phi}$ and $\boldsymbol{\psi}$ respectively, and also has minimum entropy. This is equivalent to maximizing the mutual information between the random variables having the two distributions. In earlier work [13, 14], a symmetrized version of the mutual information is used to define a metric distance, called the 'variation of information' metric, between *random variables* assuming values in distinct sets. Since our definition is an extension of this idea, we too use the same nomenclature and refer to the distance defined here (but between probability distributions and not random variables) as the variation of information metric.

We then proceed to solve the problem of actually computing the distance by finding the (or a) $\boldsymbol{\theta}$ with minimum entropy, given $\boldsymbol{\phi}, \boldsymbol{\psi}$. This is facilitated by first proving a principle of optimality, which allows us to break down the original problem into smaller and smaller problems. Using the principle of optimality, we show that in the case where $m = 2$, the optimization problem is NP-hard, because it is equivalent to a nonstandard version of the bin-packing problem. It is therefore plausible that the problem continues to be NP-hard even for $m \geq 3$, but we do not explore this issue. Instead we propose a greedy algorithm (for general values of $m$, not just $m = 2$) that provides a lower bound on the mutual information, and an upper bound on the metric distance between the two distributions. This algorithm has complexity $O((n + m^2) \log m)$.

Finally, once we have a distance measure between two probability distributions on distinct sets, it is natural to study the optimal order-reduction problem. Specifically, suppose an $n$-dimensional distribution $\boldsymbol{\phi}$ and an integer $m < n$ are specified. The problem is to find a (or the) $m$-dimensional distribution $\boldsymbol{\psi}$ such that the distance $d(\boldsymbol{\phi}, \boldsymbol{\psi})$ is minimal. It is shown that all optimal approximations must be aggregations of $\boldsymbol{\phi}$, that is, obtained by adding together components of $\boldsymbol{\phi}$. Moreover, an aggregation of $\boldsymbol{\phi}$ is an optimal approximation if and only if it has maximum entropy amongst all aggregations. It turns out that the problem of optimal aggregation is yet another bin-packing problem and thus NP-hard, so we propose yet another greedy algorithm for this problem, and also provide a bound on its performance. The complexity of this algorithm is $O(n \log m)$.

## 2 The Variation of Information Metric

### 2.1 Concepts from Information Theory

We begin with a bit of notation. Throughout the paper, we shall use the symbols $\mathbb{A}, \mathbb{B}, \mathbb{C}$ for finite sets of cardinality $n, m, l$ respectively. The symbols $X, Y, Z$ denote random variables assuming values in $\mathbb{A}, \mathbb{B}, \mathbb{C}$ respectively. The symbols $\boldsymbol{\phi}, \boldsymbol{\psi}, \boldsymbol{\xi}$ denote probability distributions on the sets $\mathbb{A}, \mathbb{B}, \mathbb{C}$ respectively. None of these symbols is used in any other way. So in particular when we write $X$, it goes without saying that its probability distribution is $\boldsymbol{\phi}$. Though the elements of these sets could be any abstract entities, to avoid notational clutter we shall write $\mathbb{A} = \{1, \ldots, n\}$ instead of the more precise $\mathbb{A} = \{a_1, \ldots, a_n\}$ etc. Let $\mathbf{e}$ denote the column vector of all one's, and the subscript denote its dimension. Thus $\mathbf{e}_n$ is a column vector of $n$ one's. A matrix $P \in [0, 1]^{m \times n}$ is said to be **stochastic** if $P\mathbf{e}_n = \mathbf{e}_m$, that is, for each row, the sum of all columns equals one. Note that $P$ need not be a square matrix; but this definition is consistent with the more familiar usage for square matrices. The set of $m \times n$ stochastic matrices is denoted by $\mathbb{S}_{m \times n}$. If we take the degenerate case of $m = 1$, then the symbol $\mathbb{S}_n = \mathbb{S}_{1 \times n}$ denotes the set of nonnegative (row) vectors that add up to one. Clearly $\mathbb{S}_n$ can be identified with the set $\mathcal{M}(\mathbb{A})$ of all probability distributions on $\mathbb{A}$.

Suppose $X, Y$ are random variables assuming values in $\mathbb{A}, \mathbb{B}$ respectively, and let $\boldsymbol{\theta} \in \mathcal{M}(\mathbb{A} \times \mathbb{B})$ denote their joint distribution. For each index $i$ between 1 and $n$, let $\mathbf{p}_i$ denote the conditional distribution of $Y$ given that $X = i$. That is

$$p_{ij} = \frac{\theta_{ij}}{\sum_{j'=1}^{m} \theta_{ij'}}.$$

If $\Pr\{X = i\} = 0$, then the conditional distribution of $Y$, conditioned on the impossible event $X = i$, is chosen as its marginal distribution $\boldsymbol{\psi}$. Note that the matrix $P = [p_{ij}]$ belongs to $\mathbb{S}_{n \times m}$, and the $i$-th row of $P$, denoted by $\mathbf{p}_i$, belongs to $\mathbb{S}_m$ for each $i$. If we represent the joint distribution of $X$ and $Y$ by an $n \times m$ matrix $\Theta = [\theta_{ij}]$ where $\theta_{ij} = \Pr\{X = i \& Y = j\}$, then we can write

$$P = [\text{Diag}(\boldsymbol{\phi})]^{-1}\Theta, \tag{1}$$

where $\text{Diag}(\boldsymbol{\phi})$ represents the $n \times n$ diagonal matrix with $\phi_1, \ldots, \phi_n$ as the diagonal elements. Suppose we now define $Q \in \mathbb{S}_{m \times n}$ by

$$q_{ji} = \Pr\{X = i | Y = j\}.$$

Then it is easy to see that the following identities hold:

$$\Theta = \text{Diag}(\boldsymbol{\phi})P = Q^T \text{Diag}(\boldsymbol{\psi}), \text{ and } Q = [\text{Diag}(\boldsymbol{\psi})]^{-1} P^T \text{Diag}(\boldsymbol{\phi}). \tag{2}$$

Now we introduce various concepts from information theory. All the concepts introduced below are discussed in [6, Chapter 2]. The function $h : [0, 1] \to \mathbb{R}_+$ is defined by $h(r) = -r \log r$, with the standard convention that $h(0) = 0$. Note that $h$ is continuously differentiable except at $r = 0$, and that $h'(r) = -(1 + \log r)$. The symbol $H$ denotes the Shannon entropy of a probability distribution. Thus if $\boldsymbol{\phi} \in \mathbb{S}_n$, then

$$H(\boldsymbol{\phi}) = -\sum_{i=1}^{n} \phi_i \log \phi_i = \sum_{i=1}^{n} h(\phi_i).$$

We shall not make any distinction between the entropy of a probability distribution, and the entropy of a random variable having that probability distribution. Thus if $X$ is a random variable assuming values in $\mathbb{A}$ with probability distribution $\boldsymbol{\phi}$, then we shall use the symbols $H(\boldsymbol{\phi})$ and $H(X)$ interchangeably.

We define the **conditional entropy** of $Y$ given $X$ as

$$H(Y|X) = \sum_{i=1}^{n} \phi_i H(\mathbf{p}_i) = \sum_{i=1}^{n} \phi_i \sum_{j=1}^{m} h(p_{ij}) = -\sum_{i=1}^{n} \phi_i \sum_{j=1}^{m} p_{ij} \log p_{ij},$$

where $\mathbf{p}_i$ denotes the $i$-th row of the matrix $P$. With this definition the identities

$$H(X, Y) = H(X) + H(Y|X) = H(Y) + H(X|Y) \tag{3}$$

hold. The **mutual information** between $X$ and $Y$ is defined as

$$I(X, Y) = H(X) + H(Y) - H(X, Y) = H(Y) - H(Y|X) = H(X) - H(X|Y). \tag{4}$$

## 2.2 Setting Up the Problem

Suppose $X, Y$ are random variables assuming values in the sets $\mathbb{A}, \mathbb{B}$ respectively, with distributions $\boldsymbol{\phi}, \boldsymbol{\psi}$ respectively. We ask: *What is the maximum possible mutual information between $X$ and $Y$?* Clearly this is equivalent to asking the question: What is a (or the) distribution $\boldsymbol{\theta}$ on $\mathbb{A} \times \mathbb{B}$ that has minimum entropy, while satisfying the boundary conditions $\boldsymbol{\theta}_{\mathbb{A}} = \boldsymbol{\phi}, \boldsymbol{\theta}_{\mathbb{B}} = \boldsymbol{\psi}$? (Here it is obvious that by $\boldsymbol{\theta}_{\mathbb{A}}$ we mean the marginal distribution of $\boldsymbol{\theta}$ on $\mathbb{A}$.) Another way of posing the question is this: Given random variables $X$ and $Y$, how close can we come to making $Y$ a deterministic function of $X$ (and vice versa) by suitably selecting their joint distribution?

**Definition 1** *Given sets $\mathbb{A}, \mathbb{B}$ with $|\mathbb{A}| = n, |\mathbb{B}| = m$, and given $\boldsymbol{\phi} \in \mathbb{S}_n, \boldsymbol{\psi} \in \mathbb{S}_m$, define*

$$W(\boldsymbol{\phi}, \boldsymbol{\psi}) := \min_{\boldsymbol{\theta} \in \mathcal{M}(\mathbb{A} \times \mathbb{B})} \{H(\boldsymbol{\theta}) : \boldsymbol{\theta}_{\mathbb{A}} = \boldsymbol{\phi}, \boldsymbol{\theta}_{\mathbb{B}} = \boldsymbol{\psi}\}, \tag{5}$$

$$V(\boldsymbol{\phi}, \boldsymbol{\psi}) := W(\boldsymbol{\phi}, \boldsymbol{\psi}) - H(\boldsymbol{\phi}). \tag{6}$$

The set of $\boldsymbol{\theta} \in \mathcal{M}(\mathbb{A} \times \mathbb{B})$ that satisfy the boundary conditions $\boldsymbol{\theta}_{\mathbb{A}} = \boldsymbol{\phi}, \boldsymbol{\theta}_{\mathbb{B}} = \boldsymbol{\psi}$ is certainly not empty, because the product distribution $\boldsymbol{\phi} \times \boldsymbol{\psi}$ satisfies this requirement. Since the feasible region of $\boldsymbol{\theta}$ is nonempty and compact, and $H(\boldsymbol{\theta})$ is a continuous function of $\boldsymbol{\theta}$, the minimum in (5) is certainly achieved, and as a result both $W(\boldsymbol{\phi}, \boldsymbol{\psi})$ and $V(\boldsymbol{\phi}, \boldsymbol{\psi})$ are well-defined. Moreover, it is obvious that

$$W(\boldsymbol{\psi}, \boldsymbol{\phi}) = W(\boldsymbol{\phi}, \boldsymbol{\psi}), V(\boldsymbol{\psi}, \boldsymbol{\phi}) = V(\boldsymbol{\phi}, \boldsymbol{\psi}) + H(\boldsymbol{\phi}) - H(\boldsymbol{\psi}), \tag{7}$$

where the second identity follows from (3). Hence $W$ is symmetric in its two arguments, whereas $V$ is not; however, $V(\boldsymbol{\psi}, \boldsymbol{\phi})$ and $V(\boldsymbol{\phi}, \boldsymbol{\psi})$ are related via (7).

Given two random variables $X, Y$ taking values in $\mathbb{A}, \mathbb{B}$ and having distributions $\boldsymbol{\phi}, \boldsymbol{\psi}$ respectively, we see that: $W(\boldsymbol{\phi}, \boldsymbol{\psi})$ is the minimum entropy of the joint random variable $(X, Y)$, $V(\boldsymbol{\phi}, \boldsymbol{\psi})$ is the minimum conditional entropy $H(Y|X)$, while $H(\boldsymbol{\psi}) - V(\boldsymbol{\phi}, \boldsymbol{\psi}) = H(\boldsymbol{\phi}) - V(\boldsymbol{\psi}, \boldsymbol{\phi})$ is the maximum mutual information between $X$ and $Y$.

## 2.3 The Variation of Information Metric

Finally we come to the definition of the metric. We begin by defining a metric between *random variables*, and then move on to distributions.

**Definition 2** *Given two random variables $X, Y$, the **variation of information** between them is defined as*

$$v(X, Y) = H(X|Y) + H(Y|X). \tag{8}$$

This measure is introduced in [13, 14] where it is referred to as the 'variation of information' metric between random variables. So we retain the same nomenclature, though our metric is between probability distributions. By making liberal use of (4), we can derive several equivalent expressions for the variation of information.

$$\begin{aligned} v(X,Y) &= H(X|Y) + H(Y|X) = H(X) + H(Y) - 2I(X,Y) \\ &= H(X,Y) - I(X,Y) = 2H(X,Y) - H(X) - H(Y). \end{aligned} \qquad (9)$$

**Theorem 1** *The function $v(\cdot,\cdot)$ satisfies the axioms of a pseudometric. Thus $v$ has the properties that for all random variables $X, Y, Z$, we have $v(X,Y) \geq 0$, $v(X,Y) = v(Y,X)$, and $v(X,Y) \leq v(X,Z) + v(Y,Z)$.*

**Proof:** It is obvious that $v(X,Y) \geq 0$, and it follows from (8) that $v(X,Y) = v(Y,X)$. To show that $v(\cdot,\cdot)$ satisfies the triangle inequality, we first prove a one-sided triangle inequality by combining basic properties of entropy and conditional entropy. A good reference for these details is [6]. Note that some of these inequalities also hold over infinite sets, but such generality is not required in the present paper. Suppose $X, Y, Z$ are random variables, assuming values in finite sets $\mathbb{A}, \mathbb{B}, \mathbb{C}$ respectively. First, by definition we have

$$H(X,Z) = H(Z) + H(X|Z).$$

This identity remains valid if everything is conditioned on $Y$. Thus

$$H(X,Z|Y) = H(Z|Y) + H((X|Z)|Y).$$

Then we have the law of iterated conditioning which states that

$$H((X|Z)|Y) = H(X|Y,Z).$$

In other words, whether we condition $X$ simultaneously on the pair $(Y,Z)$, or first condition $X$ on $Z$ and then condition the resulting $X|Z$ on $Y$, the resulting conditional entropies are the same. Next,

$$H(X|Y,Z) \leq H(X|Y).$$

In other words, conditioning on more variables cannot increase entropy. Finally,

$$H(X,Z|Y) \geq H(X|Y).$$

In other words, the entropy of the pair $H(X,Z)$ is no smaller than the entropy of the single random variable $H(X)$, and this statement remains valid even if everything is conditioned on $Y$. All of these arguments can be combined into the following chain of reasoning:

$$\begin{aligned} H(X|Y) &\leq H(X,Z|Y) = H(Z|Y) + H(X|Z,Y) \\ &\leq H(Z|Y) + H(X|Z). \end{aligned} \qquad (10)$$

To prove the triangle inequality, invoke the one-sided triangle inequality (10) and observe that

$$\begin{aligned} v(X,Y) &= H(X|Y) + H(Y|X) \\ &\leq H(X|Z) + H(Z|Y) + H(Y|Z) + H(Z|X) \\ &= v(X,Z) + v(Y,Z). \end{aligned}$$

This completes the proof. ∎

Now we turn the above pseudometric *between random variables* into a pseudometric *between probability distributions*.

**Definition 3** *Given two probability distributions $\boldsymbol{\phi} \in \mathbb{S}_n, \boldsymbol{\psi} \in \mathbb{S}_m$, the* **variation of information metric** *between them is defined as*

$$d(\boldsymbol{\phi}, \boldsymbol{\psi}) = V(\boldsymbol{\phi}, \boldsymbol{\psi}) + V(\boldsymbol{\psi}, \boldsymbol{\phi}). \tag{11}$$

An alternate expression for $d$ is a ready consequence of (6).

$$d(\boldsymbol{\phi}, \boldsymbol{\psi}) := 2W(\boldsymbol{\phi}, \boldsymbol{\psi}) - H(\boldsymbol{\phi}) - H(\boldsymbol{\psi}).$$

Also, it is clear from the definition that $d(\boldsymbol{\phi}, \boldsymbol{\psi})$ is the minimum value of the variation of information $v(X,Y)$ over all random variables $X, Y$ with marginal distributions $\boldsymbol{\phi}, \boldsymbol{\psi}$ respectively.

**Theorem 2** *The function $d$ defined in (11) is a pseudometric in that it is nonnegative, symmetric and satisfies the triangle inequality.*

**Proof:** It is obvious that $d$ is nonnegative and symmetric; so it only remains to prove the triangle inequality. To prove this, we first establish a small technical point. Suppose $\boldsymbol{\eta} \in \mathcal{M}(\mathbb{A} \times \mathbb{C}), \boldsymbol{\zeta} \in \mathcal{M}(\mathbb{B} \times \mathbb{C})$ and that $\boldsymbol{\eta}_{\mathbb{C}} = \boldsymbol{\zeta}_{\mathbb{C}} = \boldsymbol{\xi}$. Then it is always possible to find a distribution $\boldsymbol{\nu} \in \mathcal{M}(\mathbb{A} \times \mathbb{B} \times \mathbb{C})$ such that $\boldsymbol{\nu}_{\mathbb{A} \times \mathbb{C}} = \boldsymbol{\eta}$ and $\boldsymbol{\nu}_{\mathbb{B} \times \mathbb{C}} = \boldsymbol{\zeta}$. In words, the claim is that, given two joint distributions, one of $X$ and $Z$, and another of $Y$ and $Z$, both of them having the same marginal distribution for $Z$, it is possible to find a joint distribution for all three variables $X, Y, Z$ such that the marginal distributions of $(X, Y)$ and of $(Y, Z)$ match the two given joint distributions. There is no claim that such a triple distribution is unique – only that such a distribution exists. To establish the claim, we construct $\boldsymbol{\nu}$ by making $X$ and $Y$ conditionally independent given $Z$, or equivalently, by making $X \to Z \to Y$ into a very short Markov chain. Accordingly, let

$$\nu_{ijk} = \frac{\eta_{ik} \zeta_{jk}}{\xi_k}.$$

It is routine to verify that $\boldsymbol{\nu}$ has the required properties, using the identities

$$\xi_k = \sum_{i \in \mathbb{A}} \eta_{ik} = \sum_{j \in \mathbb{B}} \zeta_{jk}.$$

Now we return to the proof that $d$ satisfies the triangle inequality. Given three different probability distributions $\boldsymbol{\phi} \in \mathcal{M}(\mathbb{A})$, $\boldsymbol{\psi} \in \mathcal{M}(\mathbb{B})$, $\boldsymbol{\xi} \in \mathcal{M}(\mathbb{C})$, let us choose distributions $\boldsymbol{\theta} \in \mathcal{M}(\mathbb{A} \times \mathbb{B})$, $\boldsymbol{\eta} \in \mathcal{M}(\mathbb{A} \times \mathbb{C})$ and $\boldsymbol{\zeta} \in \mathcal{M}(\mathbb{B} \times \mathbb{C})$ such that

$$\boldsymbol{\theta}_\mathbb{A} = \boldsymbol{\phi}, \boldsymbol{\theta}_\mathbb{B} = \boldsymbol{\psi}, H(\boldsymbol{\theta}) = W(\boldsymbol{\phi}, \boldsymbol{\psi}), \tag{12}$$

$$\boldsymbol{\eta}_\mathbb{A} = \boldsymbol{\phi}, \boldsymbol{\eta}_\mathbb{C} = \boldsymbol{\xi}, H(\boldsymbol{\eta}) = W(\boldsymbol{\phi}, \boldsymbol{\xi}), \tag{13}$$

$$\boldsymbol{\zeta}_\mathbb{B} = \boldsymbol{\psi}, \boldsymbol{\zeta}_\mathbb{C} = \boldsymbol{\xi}, H(\boldsymbol{\zeta}) = W(\boldsymbol{\psi}, \boldsymbol{\xi}). \tag{14}$$

In other words, $\boldsymbol{\theta}, \boldsymbol{\eta}, \boldsymbol{\zeta}$ are chosen to be optimal for maximizing the mutual information between $X$ and $Y$, between $X$ and $Z$, and between $Y$ and $Z$ respectively. As pointed out earlier, the minimum in (5) is certainly achieved, so there is no difficulty about this. Now choose $\boldsymbol{\nu}$ to be any distribution on $\mathbb{A} \times \mathbb{B} \times \mathbb{C}$ such that

$$\boldsymbol{\nu}_{\mathbb{A} \times \mathbb{C}} = \boldsymbol{\eta}, \ \boldsymbol{\nu}_{\mathbb{B} \times \mathbb{C}} = \boldsymbol{\zeta}. \tag{15}$$

As shown above, such a choice of $\boldsymbol{\nu}$ is always possible, though certainly not unique. Depending on how we choose $\boldsymbol{\nu}$ above, there is no reason to assume that $\boldsymbol{\nu}_{\mathbb{A} \times \mathbb{B}}$ equals $\boldsymbol{\theta}$. Indeed this may not even be possible. That is precisely the point. Let $X, Y, Z$ be three random variables with the joint distribution $\boldsymbol{\nu}$. Then the triangle inequality for the quantity $v$ shows that

$$v(X, Y) \leq v(X, Z) + v(Y, Z).$$

The manner in which $\boldsymbol{\eta}$ and $\boldsymbol{\zeta}$ were chosen shows that

$$v(X, Z) = d(\boldsymbol{\phi}, \boldsymbol{\xi}), \ v(Y, Z) = d(\boldsymbol{\psi}, \boldsymbol{\xi}).$$

However, an analogous statement about $v(X, Y)$ may not be true. So we note instead that $d(\boldsymbol{\phi}, \boldsymbol{\psi})$ is the *minimum* of $v(X, Y)$ whenever $X$ and $Y$ have distributions $\boldsymbol{\phi}, \boldsymbol{\xi}$ respectively. Hence

$$d(\boldsymbol{\phi}, \boldsymbol{\psi}) \leq v(X, Y) \leq v(X, Z) + v(Y, Z) = d(\boldsymbol{\phi}, \boldsymbol{\xi}) + d(\boldsymbol{\psi}, \boldsymbol{\xi}),$$

which is the desired conclusion. ∎

## 2.4 Properties of the Metric

We conclude this section with a few simple observations.

**Theorem 3** *If $\boldsymbol{\phi} \in \mathbb{S}_n, \boldsymbol{\psi} \in \mathbb{S}_m$ and both have strictly positive elements, then $d(\boldsymbol{\phi}, \boldsymbol{\psi}) = 0$ if and only if $n = m$ and $\boldsymbol{\phi}, \boldsymbol{\psi}$ are permutations of each other.*

**Proof:** From the definitions, it is clear that $d(\boldsymbol{\phi}, \boldsymbol{\psi}) = 0$ if and only if $V(\boldsymbol{\phi}, \boldsymbol{\psi}) = V(\boldsymbol{\psi}, \boldsymbol{\phi}) = 0$. In other words, it should be possible to choose a joint distribution $\boldsymbol{\theta}$ such that $\boldsymbol{\theta}_\mathbb{A} = \boldsymbol{\phi}, \boldsymbol{\theta}_\mathbb{B} = \boldsymbol{\psi}$ and in addition, $H(X|Y) = H(Y|X) = 0$ where $X, Y$ are the random variables associated with $\boldsymbol{\phi}, \boldsymbol{\psi}$. Since both $\boldsymbol{\phi}, \boldsymbol{\psi}$ are strictly positive, this is possible only if the matrices of conditional probabilities $P$ and $Q$ (defined in (1) and its analog) consist of only zeros and ones. In turn this is possible (if and) only if $n = m$ and $\boldsymbol{\phi}, \boldsymbol{\psi}$ are permutations of each other, so that $X, Y$ are deterministic functions of each other. ∎

Theorem 3 readily implies that the metric $d$ is not a convex function. To see this, let

$$\boldsymbol{\phi} = [0.3 \ 0.7], \boldsymbol{\psi} = [0.7 \ 0.3], \boldsymbol{\mu} = 0.5\boldsymbol{\phi} + 0.5\boldsymbol{\psi} = [0.5 \ 0.5].$$

Then $d(\boldsymbol{\phi}, \boldsymbol{\phi}) = 0$, and $d(\boldsymbol{\phi}, \boldsymbol{\psi}) = 0$ because $\boldsymbol{\psi}$ is a permutation of $\boldsymbol{\phi}$. However, since $\boldsymbol{\mu}$ is *not* a permutation of $\boldsymbol{\phi}$, it follows that $d(\boldsymbol{\phi}, \boldsymbol{\mu}) > 0$.

# 3 Computing the Metric

## 3.1 Problem Formulation and Elementary Properties

Now that we have defined the metric, the next step is to compute it. For this purpose, we have one of two options. Given $\boldsymbol{\phi} \in \mathbb{S}_n, \boldsymbol{\psi} \in \mathbb{S}_m$, we can either compute the function $W(\boldsymbol{\phi}, \boldsymbol{\psi})$ by minimizing the entropy of the joint distribution, or else compute the function $V(\boldsymbol{\phi}, \boldsymbol{\psi})$ by minimizing the conditional entropy $H(Y|X)$. Note that if we compute $V(\boldsymbol{\phi}, \boldsymbol{\psi})$, then $V(\boldsymbol{\psi}, \boldsymbol{\phi})$ is automatically determined by (7). Also, minimizing the conditional entropy maximizes the mutual information, so we refer to this approach as MMI. For reasons that will become later, we assume that $n \geq m$. Clearly there is no loss of generality in doing this. The next step is to reparametrize the problem, by changing the variable of optimization from the joint distribution $\boldsymbol{\theta} \in \mathbb{S}_{nm}$ to the matrix of conditional probabilities $P \in \mathbb{S}_{n \times m}$. Thus the boundary conditions $\boldsymbol{\theta}_\mathbb{A} = \boldsymbol{\phi}, \boldsymbol{\theta}_\mathbb{B} = \boldsymbol{\psi}$ get replaced by $\boldsymbol{\phi} P = \boldsymbol{\psi}$. Also, it is clear that, for a particular choice of $P$, the conditional entropy $H(Y|X)$ is given by

$$J_{\boldsymbol{\phi}}(P) = \sum_{i=1}^{n} \phi_i H(\mathbf{p}_i), \tag{16}$$

where $\mathbf{p}_i$ is the $i$-th row of $P$. Moreover, it follows from (3) that if $P$ and $Q$ are related by (2), then

$$J_{\boldsymbol{\psi}}(Q) = J_{\boldsymbol{\phi}}(P) + H(\boldsymbol{\phi}) - H(\boldsymbol{\psi}). \tag{17}$$

Finally, it is easy to see that, given $\boldsymbol{\phi} \in \mathbb{S}_n, \boldsymbol{\psi} \in \mathbb{S}_m$, the quantity $V$ defined in (6) can also be defined equivalently as

$$V(\boldsymbol{\phi}, \boldsymbol{\psi}) = \min_{P \in \mathbb{S}_{n \times m}} J_{\boldsymbol{\phi}}(P). \tag{18}$$

Since maximizing mutual information is equivalent to minimizing conditional entropy, we can now formulate the problem under study precisely.

**MMI Problem:** Given $\boldsymbol{\phi} \in \mathbb{S}_n, \boldsymbol{\psi} \in \mathbb{S}_m$, find a $P \in \mathbb{S}_{n \times m}$ that minimizes $J_{\boldsymbol{\phi}}(P)$ subject to the boundary condition $\boldsymbol{\phi} P = \boldsymbol{\psi}$.

It is clear that the feasible region for this problem

$$\mathcal{F} := \{P \in \mathbb{S}_{n \times m} : \boldsymbol{\phi} P = \boldsymbol{\psi}\} \tag{19}$$

is a polyhedral convex set, because it is defined by a finite number of linear equalities and inequalities. Recall that an element of a convex set is said to be an **extreme point** if it cannot be expressed as a nontrivial convex combination of two other points belonging to the set. Since $\mathcal{F}$ is polyhedral, it has only a finite number of extremal points.

**Theorem 4** *Suppose all elements of $\boldsymbol{\phi}$ are strictly positive. Then the solution to the optimization problem in (17) occurs at an extreme point of $\mathcal{F}$. Thus if $P$ achieves the minimum of $J_{\boldsymbol{\phi}}(\cdot)$, then at least one element of $P$ is zero.*

**Proof:** The objective function $J_{\boldsymbol{\phi}}(\cdot)$ is strictly concave, because as is well-known, the entropy function $H(\cdot)$ is strictly concave, and $J_{\boldsymbol{\phi}}$ is just a positive linear combination of $n$ strictly concave functions. So if $P$ is not an extreme point of $\mathcal{F}$, say $P = \lambda Q + (1-\lambda) R$ where $\lambda > 0$ and $Q, R \in \mathcal{F}$ with $Q \neq R$, then the strict concavity of $J_{\boldsymbol{\phi}}$ implies that

$$J_{\boldsymbol{\phi}}(P) = J_{\boldsymbol{\phi}}(\lambda Q + (1-\lambda)R) > \lambda J_{\boldsymbol{\phi}}(Q) + (1-\lambda) J_{\boldsymbol{\phi}}(R).$$

So at least one of $J_{\boldsymbol{\phi}}(Q), J_{\boldsymbol{\phi}}(R)$ must be less than $J_{\boldsymbol{\phi}}(P)$. Clearly any matrix $P$ with all positive elements is in the interior of $\mathbb{S}_{n \times m}$ and is thus in the (relative) interior of $\mathcal{F}$ and thus is not extreme, and thus cannot be an optimum.

## 3.2 A Principle of Optimality

Given the additive nature of the objective function $J_{\boldsymbol{\phi}}(P)$, it is hardly surprising that we can prove a principle of optimality for this problem. In spite of its simplicity, the principle of optimality is extremely useful, in that it permits us to reduce large problems into a succession of smaller problems.

To state the principle of optimality, we introduce a little bit of notation. Suppose $\boldsymbol{\phi} \in \mathbb{S}_n, \boldsymbol{\psi} \in \mathbb{S}_m$ are specified, and that $\phi_i > 0$ for all $i$. Suppose $\mathbb{A} = \{1, \ldots, n\}$, and let $\mathbb{A}'$ be a nonempty proper subset of $\mathbb{A}$. For notational convenience, suppose $\mathbb{A}' = \{1, \ldots, k\}$ where $k < n$. For $\boldsymbol{\phi} \in \mathbb{S}_n, P \in \mathbb{S}_{n \times m}$, define

$$\boldsymbol{\phi}' := [\phi_1 \ldots \phi_k], P' := \begin{bmatrix} \mathbf{p}_1 \\ \vdots \\ \mathbf{p}_k \end{bmatrix},$$

and note that $P' \in \mathbb{S}_{k \times m}$, though in general $\boldsymbol{\phi}'$ need not belong to $\mathbb{S}_k$. After this elaborate build-up we can now state the principle of optimality.

**Theorem 5** *With all notation as above, suppose $\phi_i > 0\ \forall i$, and suppose that $P^*$ minimizes $J_{\boldsymbol{\phi}}(P)$ subject to the constraint that $\boldsymbol{\phi} P = \boldsymbol{\psi}$. Define $c = \boldsymbol{\phi}' \mathbf{e}_k > 0$, and $\boldsymbol{\psi}' = \sum_{i=1}^{k} \phi_i \mathbf{p}_i^* = \boldsymbol{\phi}'(P^*)'$. Observe that $(1/c)\boldsymbol{\phi}' \in \mathbb{S}_k, (1/c)\boldsymbol{\psi}' \in \mathbb{S}_m$. Then $(P^*)'$ minimizes $J_{\boldsymbol{\phi}'}(P')$ over $\mathbb{S}_{k \times m}$ subject to the constraint that $(1/c)\boldsymbol{\phi}'P' = (1/c)\boldsymbol{\psi}'$.*

**Proof:** Note that $(P^*)'$ is also a stochastic matrix in that $(P^*)'\mathbf{e}_m = \mathbf{e}_k$. Hence

$$\boldsymbol{\psi}'\mathbf{e}_m = \boldsymbol{\phi}'(P^*)'\mathbf{e}_m = \boldsymbol{\phi}'\mathbf{e}_k = c > 0,$$

because every component of $\boldsymbol{\phi}$ is positive. Hence $\boldsymbol{\psi}'$ is certainly not the zero vector, even though some components of $\boldsymbol{\psi}'$ could be zero. Thus $(1/c)\boldsymbol{\phi}' \in \mathbb{S}_k, (1/c)\boldsymbol{\psi}' \in \mathbb{S}_m$, and the minimization problem under study is similar to the larger problem. To prove the claim, suppose by way of contradiction that there exists another matrix $Q' \in \mathbb{S}_{k \times m}$ such that

$$\boldsymbol{\phi}'Q' = \boldsymbol{\psi}', J_{\boldsymbol{\phi}'}(Q') = \sum_{i=1}^{k} \phi_i H(\mathbf{q}_i) < J_{\boldsymbol{\phi}'}((P^*)') = \sum_{i=1}^{k} \phi_i H(\mathbf{p}_i^*).$$

Define in an analogous fashion

$$(P^*)'' = \begin{bmatrix} \mathbf{p}_{k+1} \\ \vdots \\ \mathbf{p}_n \end{bmatrix}, Q = \begin{bmatrix} Q' \\ (P^*)'' \end{bmatrix},$$

and note that, since $P^*$ is feasible for the original problem, we have that

$$\sum_{i=k+1}^{n} \phi_i \mathbf{p}_i^* = \boldsymbol{\phi} P^* - \boldsymbol{\phi}'(P^*)' = \boldsymbol{\psi} - \boldsymbol{\psi}'.$$

Now

$$J_\phi(Q) = \sum_{i=1}^{k} \phi_i H(\mathbf{q}_i) + \sum_{i=k+1}^{n} \phi_i H(\mathbf{p}_i^*)$$

$$< \sum_{i=1}^{k} \phi_i H(\mathbf{p}_i^*) + \sum_{i=k+1}^{n} \phi_i H(\mathbf{p}_i^*) = J_\phi(P^*),$$

while

$$\phi Q = \phi' Q' + \sum_{i=k+1}^{n} \phi_i \mathbf{p}_i^* = \psi' + \psi - \psi' = \psi.$$

Hence $Q$ is feasible for the original problem and has a lower objective function, which is a contradiction. Hence $(P^*)'$ is a minimizer of the reduced-size problem. ∎

## 4 Solution to the MMI Problem in the $n \times 2$ Case

### 4.1 The $2 \times 2$ Case

In this subsection we give an explicit closed-form expression for $V(\phi, \psi)$ when $n = m = 2$ and $\phi, \psi \in \mathbb{S}_2$. Without loss of generality, assume that $\phi \neq \psi$ because $V(\phi, \psi) = 0$ if $\phi = \psi$. Also, again without loss of generality, rearrange the elements of $\phi, \psi$ such that both vectors are in strictly increasing order.[3] Then we can distinguish between two cases, namely (a) $0 < \psi_1 < \phi_1 < \phi_2 < \psi_2$, and (b) $0 < \phi_1 < \psi_1 < \psi_2 < \phi_2$.

**Theorem 6** *Suppose $n = m = 2$ and $\phi, \psi \in \mathbb{S}_2$. Suppose further that we have either case (a) or case (b) above. If $0 < \psi_1 < \phi_1 < \phi_2 < \psi_2$, then*

$$V(\phi, \psi) = \phi_1 H(\mathbf{v}_{21}) = -\phi_1 \left[ (\psi_1/\phi_1) \log(\psi_1/\phi_1) + ((1 - \psi_1/\phi_1) \log(1 - \psi_1/\phi_1) \right], \quad (20)$$

*where*

$$\mathbf{v}_{21} = \begin{bmatrix} \frac{\psi_1}{\phi_1} & 1 - \frac{\psi_1}{\phi_1} \end{bmatrix}.$$

*If $0 < \phi_1 < \psi_1 < \psi_2 < \phi_2$ then*

$$V(\phi, \psi) = \phi_2 H(\mathbf{v}_{12}) = -\phi_2 \left[ (1 - \psi_2/\phi_2) \log(1 - \psi_2/\phi_2) + (\psi_2/\phi_2) \log(\psi_2/\phi_2) \right], \quad (21)$$

*where*

$$\mathbf{v}_{12} = \begin{bmatrix} 1 - \frac{\psi_2}{\phi_2} & \frac{\psi_2}{\phi_2} \end{bmatrix}.$$

---

[3] To avoid unnecessary pedantry, we assume that lots of strict inequalities hold. The modifications needed to handle the case where some of the inequalities are not strict are easy and are left to the reader.

**Proof:** From Theorem 4, we know that any optimal choice of $P \in \mathbb{S}_{2\times 2}$ must be an extreme point of the feasible region. Thus at least one component of $P$ must be zero. The constraints that $P$ is stochastic and that $\boldsymbol{\phi} P = \boldsymbol{\psi}$ lead to the following four possible extreme points of the feasible region.

$$P_{11} = \begin{bmatrix} 0 & 1 \\ \frac{\psi_1}{\phi_2} & 1 - \frac{\psi_1}{\phi_2} \end{bmatrix}, P_{12} = \begin{bmatrix} 1 & 0 \\ 1 - \frac{\psi_2}{\phi_2} & \frac{\psi_2}{\phi_2} \end{bmatrix},$$

$$P_{21} = \begin{bmatrix} \frac{\psi_1}{\phi_1} & 1 - \frac{\psi_1}{\phi_1} \\ 0 & 1 \end{bmatrix}, P_{22} = \begin{bmatrix} 1 - \frac{\psi_2}{\phi_1} & \frac{\psi_2}{\phi_1} \\ 1 & 0 \end{bmatrix}.$$

Moreover, since $\psi_2 > \phi_2$ and $\psi_2 > \phi_1$, it follows that $P_{12}$ and $P_{22}$ are infeasible, and the only possibilities are $P_{11}$ and $P_{21}$. So all we need to do is to compute $J_{\boldsymbol{\phi}}(P_{11}), J_{\boldsymbol{\phi}}(P_{21})$, and pick the one that is smaller.

Accordingly, let us define

$$V_{11} = J_{\boldsymbol{\phi}}(P_{11}) = \phi_2[h(\psi_1/\phi_2) + h(1 - (\psi_1/\phi_2))],$$

$$V_{21} = J_{\boldsymbol{\phi}}(P_{21}) = \phi_1[h(\psi_1/\phi_1) + h(1 - (\psi_1/\phi_1))].$$

Equation (20) is established if it can be shown that $V_{21} < V_{11}$. This is just an exercise in elementary calculus. Let us replace $\psi_1$ by $u$, and define

$$f_u(\phi) := \phi[h(u/\phi) + h(1 - (u/\phi))]. \tag{22}$$

The desired conclusion follows if it can be shown that for a fixed $u > 0$, the function $\phi \mapsto f_u(\phi)$ is strictly increasing when $\phi > u$. Note that

$$\begin{aligned}
f_u(\phi) &= \phi \left[ \frac{u}{\phi} \log \frac{\phi}{u} + \frac{\phi - u}{\phi} \log \frac{\phi}{\phi - u} \right] \\
&= u \log \phi - u \log u + (\phi - u) \log \phi - (\phi - u) \log(\phi - u) \\
&= \phi \log \phi - (\phi - u) \log(\phi - u) - u \log u \\
&= -h(\phi) + h(\phi - u) + h(u).
\end{aligned}$$

Note that $h'(x) = -(1 + \log x)$. So

$$\frac{df_u}{d\phi} = 1 + \log \phi - 1 - \log(\phi - u) = \log \frac{\phi}{\phi - u} > 0 \text{ if } \phi > u.$$

So $V_{11} = f_{\psi_1}(\phi_2) > V_{22} = f_{\psi_1}(\phi_1)$ if $\phi_2 > \phi_1 > \psi_1$.

Now suppose we are in case (b). Since the optimal choice of $\boldsymbol{\theta}$ is symmetric in $\boldsymbol{\phi}$ and $\boldsymbol{\psi}$, we can apply the above formula with the roles of $\boldsymbol{\phi}$ and $\boldsymbol{\psi}$ interchanged (which would put it in case (a)). Then we interchange the symbols of $\boldsymbol{\phi}, \boldsymbol{\psi}$ and permute rows and columns in a symmetric fashion. This leads to the conclusion that, in case (b), the optimal choice of $\boldsymbol{\theta}$ is

$$\boldsymbol{\theta} = \begin{bmatrix} \phi_1 & 0 \\ \phi_2 - \psi_2 & \phi_2 \end{bmatrix},$$

and leads to the formula (21). ∎

## 4.2 The $n \times 2$ Case

We begin with a notion that is encountered again several times in the paper.

**Definition 4** *Given $\boldsymbol{\phi} \in \mathbb{S}_n, \boldsymbol{\psi} \in \mathbb{S}_m$ with $n > m$, $\boldsymbol{\psi}$ is said to be an **aggregation** of $\boldsymbol{\phi}$ if there exists a partition of $\mathbb{A}$ into $m$ sets $I_1, \ldots, I_m$ such that $\sum_{i \in I_j} \phi_i = \psi_j$ for $j = 1, \ldots, m$.*

Next we introduce a nonstandard bin-packing problem which is also encountered several times later on. The standard bin-packing problem with bin capacity of one is as follows: Given a list $a_1, \ldots, a_n$ where $a_i \leq 1 \ \forall i$, the objective is to find the smallest integer $l$ for which there exists a partition of $\{1, \ldots, n\}$ into $l$ sets $I_1, \ldots, I_l$ such that $\sum_{i \in I_j} a_i \leq 1$ for $j = 1, \ldots, l$. Thus the objective is to pack the given list into as few bins as possible. This problem is known to be NP-hard. Now consider a variant of the problem that is relevant to the present situation. Given $\boldsymbol{\phi} \in \mathbb{S}_n, \boldsymbol{\psi} \in \mathbb{S}_m$ with $n > m$, think of $\psi_1, \ldots, \psi_m$ as the capacities of $m$ bins, and of $\phi_1, \ldots, \phi_n$ as the list to be packed into these $m$ bins. Since $\sum_i \phi_i = \sum_j \psi_j = 1$, there are only two possibilities: First, $\boldsymbol{\psi}$ is an aggregation of $\boldsymbol{\phi}$, in which case all bins can be filled up exactly. Second, $\boldsymbol{\psi}$ is not an aggregation of $\boldsymbol{\phi}$, in which case some bins are overstuffed, while the rest have unutilized capacity. The bin-packing problem with overstuffing and variable bin capacities is as follows: Find a partition of $\mathbb{A}$ into $m$ sets $I_1, \ldots, I_m$ such that the total mismatch

$$MI = \sum_{j \in \mathbb{B}} \left| \psi_j - \sum_{i \in I_j} \phi_i \right|$$

is as small as possible. Note that we can also define the total unutilized capacity and the total overstuffing as

$$UC = \sum_{j \in \mathbb{B}} \left( \psi_j - \sum_{i \in I_j} \phi_i \right)_+ , OS = - \sum_{j \in \mathbb{B}} \left( \psi_j - \sum_{i \in I_j} \phi_i \right)_- .$$

In these definitions we use the standard convention that $(x)_+ = \max\{x, 0\}$ and $(x)_- = \min\{x, 0\}$. Then it is clear that $OS = UC$ and that $MI = 2OS = 2UC$. So the problems of minimizing the total mismatch, or total overstuffing, or total unutilized capacity, are all equivalent. Unfortunately, this problem is also NP-hard [7]. Even determining whether a given $\boldsymbol{\psi}$ is an aggregation of a given $\boldsymbol{\phi}$ or not is also NP-hard. The bin packing with overstuffing is discussed in [7, 3, 4] among other papers.

With this background, we now present a partial solution to the problem of computing $V(\boldsymbol{\phi}, \boldsymbol{\psi})$ when $m = 2$ in terms of the bin-packing problem with overstuffing with two bins. If $\boldsymbol{\psi}$ is an aggregation of $\boldsymbol{\phi}$, then obviously $V(\boldsymbol{\phi}, \boldsymbol{\psi}) = 0$. Otherwise, let $\psi_1, \psi_2$ denote the capacity of the two bins, and let $\phi_1, \ldots, \phi_n$ denote the list to be packed. Without loss of generality, assume that the $\phi_i$ are in decreasing order of magnitude. Let $\mathcal{N}_1, \mathcal{N}_2$ denote an optimal partition of $\mathcal{N} = \{1, \ldots, n\}$ and let $c$ denote the minimum unutilized capacity. Again, without loss of generality, assume that bin 1 is underutilized and that bin 2 is overstuffed. This means that

$$\psi_2 - \sum_{i \in \mathcal{N}_2} \phi_i = -\psi_1 + \sum_{i \in \mathcal{N}_1} \phi_i = c. \tag{23}$$

**Theorem 7** *Suppose $\boldsymbol{\psi}$ is not an aggregation of $\boldsymbol{\phi}$, and solve the bin-packing problem as above. If $n \in \mathcal{N}_2$, then an optimal choice of $P$ that minimizes $J_{\boldsymbol{\phi}}(P)$ subject to $\boldsymbol{\phi} P = \boldsymbol{\psi}$ is given by*

$$\mathbf{p}_i = [1 \ 0] \ \forall i \in \mathcal{N}_1, \mathbf{p}_i = [0 \ 1] \ \forall i \in \mathcal{N}_2 \setminus \{n\}, \mathbf{p}_n = [\ c/\phi_n \ \ (\phi_n - c)/\phi_n \ ]. \tag{24}$$

*Moreover*

$$V(\boldsymbol{\phi}, \boldsymbol{\psi}) = \phi_n H(\mathbf{p}_n) = f_c(\phi_n),$$

*where the function $f$ is defined in (22).*

**Proof:** From the principle of optimality, we know that if a matrix $P$ is optimal for the $n \times 2$ problem, then every $2 \times 2$ submatrix is optimal for its respective problem, and thus has at most one strictly positive row. Taken together this shows that any optimal choice of $P$ has at most one strictly positive row, while the rest are either $[1 \ 0]$ or $[0 \ 1]$. Accordingly, define $P$ as above, and let $R$ be another matrix that has exactly one strictly positive row such that $\boldsymbol{\phi} R = \boldsymbol{\psi}$. All we need to do is to show that $J_{\boldsymbol{\phi}}(R) \geq J_{\boldsymbol{\phi}}(P)$. For this purpose, suppose the $k$-th row of $R$ is strictly positive, and define

$$I_1 = \{i : \mathbf{r}_i = [1 \ 0]\}, I_2 = \{i : \mathbf{r}_i = [0 \ 1]\},$$

while $\mathbf{r}_k$ is strictly positive. Then $\boldsymbol{\phi} R = \boldsymbol{\psi}$ implies that

$$u_1 := \psi_1 - \sum_{i \in I_1} \phi_i > 0, u_2 := \psi_2 - \sum_{i \in I_2} \phi_i > 0, u_1 + u_2 = \phi_k,$$

$$\mathbf{r}_k = [\ u_1/\phi_k \quad u_2/\phi_k\ ], J_\phi(R) = \phi_k H(\mathbf{r}_k) = f_{u_1}(\phi_k),$$

where the function $f$ is defined in (22), and we use the fact that $u_2 = \phi_k - u_1$. The fact that $c$ is the optimal unutilized capacity implies that $c \leq \min\{u_1, u_2\}$. This is because if we add $\phi_k$ to bin 1, then bin 2 has an unutilized capacity of $u_2$, whence $u_2 \geq c$, and similarly $u_1 \geq c$. To put it another way, we have that $c \leq \min\{u_1, u_2\} \leq \max\{u_1, u_2\} \leq \phi_k - c$. In turn this implies that

$$H(\mathbf{r}_k) = H([\ u_1/\phi_k \quad u_2/\phi_k\ ]) \geq H([\ c/\phi_k \quad (\phi_k - c)/\phi_k\ ]).$$

So we now conclude that

$$\begin{aligned} J_\phi(R) &= \phi_k H(\mathbf{r}_k) \geq \phi_k H([\ c/\phi_k \quad (\phi_k - c)/\phi_k\ ]) \\ &= f_c(\phi_k) \geq f_c(\phi_n) = J_\phi(P) \end{aligned}$$

because $\phi_n \leq \phi_k$ and $f_c(\cdot)$ is a strictly increasing function. ∎

**Example 1** *Theorem 7 applies only when the last index $n$ belongs to the overstuffed bin. In case $n$ belongs to the understuffed bin, one might be tempted to modify the matrix $P$ of (24) as follows: Let $k^*$ be the smallest index in the set $\mathcal{N}_2$, and let*

$$\mathbf{p}_i = [1\ 0]\ \forall i \in \mathcal{N}_1, \mathbf{p}_i = [0\ 1]\ \forall i \in \mathcal{N}_2 \setminus \{k^*\}, \mathbf{p}_{k^*} = [\ c/\phi_{k^*} \quad (\phi_{k^*} - c)/\phi_{k^*}\ ].$$

*Unfortunately such a choice of $P$ is not always optimal, as shown here. Let $n = 5$, and*

$$\boldsymbol{\psi} = [\ 0.4 \quad 0.6\ ], \boldsymbol{\phi} = [\ 0.50 \quad 0.24 \quad 0.12 \quad 0.071 \quad 0.069\ ].$$

*Then the optimal packing is given by*

$$\mathcal{N}_1 = \{2, 4, 5\}, \mathcal{N}_2 = \{1, 3\}, c = 0.02,$$

*and the smallest element in $\mathcal{N}_2$ is $\phi_3 = 0.12$. Accordingly let us define $k^* = 3$, and*

$$P = \begin{bmatrix} 0 & 1 & 0.02/0.12 & 1 & 1 \\ 1 & 0 & 0.10/0.12 & 0 & 0 \end{bmatrix}^t.$$

*Then (using the natural logarithm to compute entropy)*

$$H(\mathbf{p}_3) = 0.4506, J_\phi(P) = \phi_3 H(\mathbf{p}_3) = 5.4067 \times 10^{-2}.$$

*If we now swap $\phi_3$ into bin 1 and $\phi_4, \phi_5$ into bin 2 so that*

$$\mathcal{N}'_1 = \{2, 3\}, \mathcal{N}'_2 = \{1, 4, 5\},$$

*then bin 2 is overstuffed by $0.04 > 0.02$ as expected. Since $\phi_5$ is now the smallest element in bin 2, let us define*

$$P' = \begin{bmatrix} 0 & 1 & 1 & 0 & 0.040/0.069 \\ 1 & 0 & 0 & 1 & 0.029/0.069 \end{bmatrix}^t.$$

*Then*

$$H(\mathbf{p}'_5) = 0.6804, \ J_\phi(P') = \phi_5 H(\mathbf{p}'_5) = 4.6947 \times 10^{-2} < J_\phi(P).$$

*So $P$ is not an optimal choice.*

## 5  Solution to the MMI Problem in the $n \times m$ Case

### 5.1  Greedy Algorithm for the MMI Problem

In general, determining whether $\psi$ is an aggregation of $\phi$, or finding the optimal bin allocations allowing overstuffing, are both NP-hard problems [3, 4]. It follows that computing $V(\phi, \psi)$, or equivalently, computing the maximum mutual information, is also NP-hard when $m = 2$. It is therefore plausible that the problem of computing $V(\phi, \psi)$ continues to be NP-hard if $3 \leq m \leq n$. But we do not explore this issue further. Instead, we borrow a standard greedy algorithm for bin-packing with overstuffing from the computer science literature [19], known as 'best fit,' and adapt it to the current situation. We begin by arranging the elements of $\psi$ in descending order. In general it is *not* necessary to sort the elements of $\phi$.

Given $\phi \in \mathbb{S}_n, \psi \in \mathbb{S}_m$ with $m < n$, proceed as follows:

1. Set $s = 1$, where $s$ is the round counter. Define $n_s = n, m_s = m, \phi_s = \phi, \psi_s = \psi$.

2. Place each element of $\phi$ in the bin with the largest unused capacity. If a particular component $(\phi_s)_i$ does not fit into any bin, assign the index $i$ to an overflow index set $K_s$.

3. When all elements of $\phi_s$ have been processed, let $I_1^{(s)}, \ldots, I_{m_s}^{(s)}$ be the indices from $\{1, \ldots, n_s\}$ that have been assigned to the various bins, and let $K_s$ denote the set of indices that cannot be assigned to any bin. If $|K_s| > 1$ go to Step 4; otherwise go to Step 5.

4. Define $\alpha_1^{(s)}, \ldots, \alpha_{m_s}^{(s)}$ to be the unutilized capacities of the $m_s$ bins, and define $\boldsymbol{\alpha}^{(s)} = [\alpha_1^{(s)} \ldots \alpha_{m_s}^{(s)}]$. Then the total unutilized capacity $c_s := \boldsymbol{\alpha}^{(s)} \mathbf{e}_{m_s}$ satisfies

$$c_s = \sum_{j=1}^{m_s} \alpha_j^{(s)} = \sum_{i \in K_s} (\phi_s)_i. \tag{25}$$

Since each $(\phi_s)_i, i \in K_s$ does not fit into any bin, it is clear that $(\phi_s)_i > \alpha_j^{(s)}, \forall i, j$. In turn this implies that $|K_s| < m_s$. Next, set $n_{s+1} = m_s, m_{s+1} = |K_s|$, and define

$$\phi_{s+1} = \frac{1}{c_s}\alpha^{(s)} \in \mathbb{S}_{n_{s+1}}, \psi_{s+1} = \frac{1}{c_s}[(\phi_s)_i] \in \mathbb{S}_{m_{s+1}}.$$

Increment the counter and go to Step 2.

5. When this step is reached, $|K_s|$ is either zero or one. If $|K_s| = 0$, then it means that $\psi_s$ is a perfect aggregation of $\phi_s$. So define $V_s = 0$ and proceed as below. If $|K_s| = 1$, then only one element of $\phi_s$, call it $(\phi_s)_k$, cannot be packed into any bin, and this component must equal $c_s$. So let

$$\mathbf{v}_s = \frac{1}{c_s}\alpha^{(s)} \in \mathbb{S}_{m_s}, V_s = c_s H(\mathbf{v}_s), U_s = V_s + H(\phi_s) - H(\psi_s).$$

Define $P_s \in \mathbb{S}_{n_s \times m_s}$ by

$$\mathbf{p}_i = \mathbf{b}_j \text{ if } i \in I_j^{(s)}, \mathbf{p}_k = \mathbf{v}_s,$$

where $\mathbf{b}_j$ is the $j$-th unit vector with $m_s$ components. Then $V_s$ is the minimum value of $J_{\phi_s}(\cdot)$, and $P_s$ achieves that minimum. Next, define $Q_s \in \mathbb{S}_{m_s \times n_s}$ by

$$Q_s = [\text{diag}(\psi_s)]^{-1} P_s^T \text{Diag}(\phi_s).$$

Then it follows from (17) that $Q_s$ minimizes $J_{\psi_s}(\cdot)$, and that $U_s$ is the value of that minimum.

6. In this step, we invert all of the above steps by transposing $Q_{s+1}$, applying the transformation in (2), and embedding the resulting matrix into $P_s$. We also correct the cost function using (17). Decrement the counter $s$ and recall that $m_s = n_{s+1}$. Recall the unutilized capacity $c_s$ defined in (25) which has been found during the forward iteration, and define

$$V_s = c_s U_{s+1}, U_s = V_s + H(\phi_s) - H(\psi_s).$$

Define $P_s \in \mathbb{S}_{n_s \times m_s}$ by

$$\mathbf{p}_i = \mathbf{b}_j \text{ if } i \in I_j^{(s)}, \mathbf{p}_i = i\text{-th row of } Q_{s+1}.$$

If $s = 1$, halt; otherwise repeat the step.

The computational complexity of algorithm is easy to bound. The first step is to sort the elements of $\psi$, which has complexity $O(m^2)$ if we insist on an exact answer or $O(m \log m)$ if we use a randomized algorithm like quick sort. We use the latter bound here. In each step of the best fit algorithm, the bin in which the current element of $\boldsymbol{\phi}$ has been placed has maximum capacity *before* placing, but necessarily *after* placing. So it needs to moved into the right place. Since the rest of the bins are still in descending order of capacity, this can be achieved in $O(\log m)$ steps using a bisection search. And this has to be done $n$ times. So once $\boldsymbol{\psi}$ is sorted, one run of the best fit algorithm has complexity $O(n \log m)$, which dominates the complexity $O(m \log m)$ of sorting $\boldsymbol{\psi}$, since $m \leq n$. Since the size of the problem decreases at each round, at worst we may have to run the best fit algorithm $m - 1$ times. Moreover, after the first round, the size of the problem is not any larger than $m \times (m - 1)$. So the overall complexity of the greedy algorithm is no worse than $O(n \log m) + mO(m \log m) = O((n + m^2) \log m)$. The fact that the complexity is only linear in $n$ is heartening.

## 5.2 Illustrative Example

The application of the greedy algorithm is illustrated via a large example that needs to go through three rounds.

**Example 2** *To illustrate the above algorithm, we solve a $40 \times 10$ problem.[4] First two uniformly distributed random vectors $\mathbf{x} \in [0,1]^{40}, \mathbf{y} \in [0,1]^{10}$ were generated using the* `rand` *command of* `Matlab`. *Then these were stretched out via the transformation*

$$\phi_i = \exp(x_i)/s_1, \psi_j = \exp(y_j)/s_2,$$

*where $s_1, s_2$ are scaling constants to make the sums come out equal to one. Then only the smaller vector is sorted in descending order. The results are shown below. For display purposes the resulting $\boldsymbol{\phi}$ is shown as a matrix, though of course it is a row vector.*

$$\boldsymbol{\phi} = \begin{bmatrix} 0.0304 & 0.0333 & 0.0153 & 0.0335 & 0.0253 & 0.0148 & 0.0178 & 0.0232 \\ 0.0350 & 0.0353 & 0.0157 & 0.0355 & 0.0350 & 0.0219 & 0.0299 & 0.0155 \\ 0.0205 & 0.0336 & 0.0297 & 0.0351 & 0.0259 & 0.0139 & 0.0314 & 0.0342 \\ 0.0265 & 0.0287 & 0.0283 & 0.0199 & 0.0259 & 0.0160 & 0.0273 & 0.0139 \\ 0.0177 & 0.0141 & 0.0148 & 0.0307 & 0.0270 & 0.0185 & 0.0348 & 0.0139 \end{bmatrix},$$

$$\boldsymbol{\psi} = [\, 0.1241 \;\; 0.1205 \;\; 0.1192 \;\; 0.1139 \;\; 0.1069 \;\; 0.0914 \;\; 0.0875 \;\; 0.0869 \;\; 0.0821 \;\; 0.0675 \,].$$

*By applying the best fit algorithm in round one, we find that*

$$I_1^{(1)} = \{1, 7, 16, 20, 32\}, I_2^{(1)} = \{2, 11, 17, 25, 33\}, I_3^{(1)} = \{3, 6, 9, 22, 28\},$$

---
[4]The diary of the example is available upon request from the author.

$$I_4^{(1)} = \{4, 15, 26, 34\}, I_5^{(1)} = \{5, 14, 21, 30\}, I_6^{(1)} = \{8, 18, 31\}, I_7^{(1)} = \{10, 23, 35, 38\},$$

$$I_8^{(1)} = \{12, 24\}, I_9^{(1)} = \{13, 27, 40\}, I_{10}^{(1)} = \{19, 29\}, K_1 = \{36, 37, 39\}.$$

The unallocated capacity $c_1 = 0.0924$.

For round 2, we therefore have

$$\boldsymbol{\phi}_2 = [\ 0.1237\ \ 0.0721\ \ 0.0183\ \ 0.0825\ \ 0.1934\ \ 0.0793\ \ 0.0639\ \ 0.1856\ \ 0.0521\ \ 0.1291\ ],$$

and

$$\boldsymbol{\psi}_2 = \frac{1}{c_1}[\ \phi(36)\ \ \phi(37)\ \ \phi(39)\ ] = [\ 0.3317\ \ 0.2917\ \ 0.3766\ ].$$

Applying the best fit algorithm to this problem results in

$$I_1^{(2)} = \{2, 5\}, I_2^{(2)} = \{3, 4, 7\}, I_3^{(2)} = \{1, 6, 9\}, K_2 = \{8, 10\}.$$

The unutilized capacity $c_2 = 0.3146$, and

$$\boldsymbol{\phi}_3 = [\ 0.2103\ \ 0.4037\ \ 0.3860\ ].$$

$$\boldsymbol{\psi}_3 = \frac{1}{c_2}[\ \phi_2(8)\ \ \phi_2(10)\ ] = [\ 0.5898\ \ 0.4102\ ],$$

For this simple problem an exact solution can be computed using Theorem 7, and is

$$P_3 = \begin{bmatrix} 0.9691 & 0.0309 \\ 0 & 1.0000 \\ 1.0000 & 0 \end{bmatrix}.$$

The matrix $Q_3$ is now computed as $[\text{Diag}(\boldsymbol{\psi}_3)]^{-1} P_3^T \text{Diag}(\boldsymbol{\phi}_3)$ and equals

$$Q_3 = \begin{bmatrix} 0.3456 & 0 & 0.6544 \\ 0.0158 & 0.9842 & 0 \end{bmatrix}.$$

As per the algorithm, we have

$$V_3 = V(\boldsymbol{\phi}_3, \boldsymbol{\psi}_3) = J_{\boldsymbol{\phi}_3}(P_3) = (\boldsymbol{\phi}_3)_1 H((P_3)_1) = 0.0290,$$

$$U_3 = V_3 + H(\boldsymbol{\phi}_3) - H(\boldsymbol{\psi}_3) = 0.4136,$$

Now we return to round 2. The rows of $Q_3$ constitute rows 8 and 10 of the $10 \times 3$ matrix $P_2$, while all other rows are elementary row vectors. The i-th row of $P_2$ equals the j-th elementary row vector if the index $i$ belongs to the set $I_j^{(2)}$. Further,

$$V_2 = J_{\boldsymbol{\phi}_2}(P_2) = c_2 U_3 = 0.1301,$$

$$U_2 = V_2 + H(\boldsymbol{\phi}_2) - H(\boldsymbol{\psi}_2) = 0.1301 + 2.1525 - 1.0932 = 1.1894.$$

At last we can get back to the initial round. From the $10 \times 3$ matrix $P_2$, we generate a matrix $Q_2 \in [0,1]^{3 \times 10}$ by the familiar transformation, which leads to

$$\begin{bmatrix} 0 & 0.2174 & 0 & 0 & 0.5831 & 0 & 0 & 0.1933 & 0 & 0.0062 \\ 0 & 0 & 0.0627 & 0.2827 & 0 & 0 & 0.2191 & 0 & 0 & 0.4354 \\ 0.3286 & 0 & 0 & 0 & 0 & 0.2105 & 0 & 0.3225 & 0.1384 & 0 \end{bmatrix}.$$

The three rows of $Q_2$ form rows 36, 37 and 39 of the matrix $P = P_1$, while the other 37 rows are elementary vectors.

Finally we compute the values of $V, U$.

$$V_1 = J_{\boldsymbol{\phi}}(P) = c_1 U_2 = 0.1099,$$

$$U_1 = V_1 + H(\boldsymbol{\phi}_1) - H(\boldsymbol{\psi}_1) = 0.1099 + 3.6399 - 2.2853 = 1.4645.$$

What this means is that, with the choice of $P$ as described above, whenever $X$ assumes any value other than 36, 37 or 39, $Y$ is a deterministic function of $X$. With this choice of $P$, we have $J_{\boldsymbol{\phi}}(P) = 0.1099$ as the conditional entropy $H(Y|X)$, a reduction of more than 95% from $H(Y) = H(\boldsymbol{\psi}) = 2.2853$. Similarly $H(X|Y) = U_1 = 1.4645$. So we conclude that

$$d(\boldsymbol{\phi}, \boldsymbol{\psi}) \leq V_1 + U_1 = 1.5744.$$

## 6 Optimal Order Reduction

Now that we have a way of measuring the metric distance between two probability distributions defined on sets of different cardinality, we can examine the problem of approximating a distribution $\boldsymbol{\phi} \in \mathbb{S}_n$ by another $\boldsymbol{\psi} \in \mathbb{S}_m$ where $m < n$, such that the distance $d(\boldsymbol{\phi}, \boldsymbol{\psi})$ between them is as small as possible. We begin by showing that all optimal reduced-order approximations must be aggregations of $\boldsymbol{\phi}$. Then we show that minimizing the distance $d(\boldsymbol{\phi}, \boldsymbol{\psi})$ is equivalent to maximizing the entropy of the aggregation $\boldsymbol{\psi}$. Then the problem of maximizing the entropy of the aggregation is formulated as yet another bin-packing problem, this time with equal-sized bins. This problem is solved via a best fit greedy algorithm, and an upper bound is presented for the performance of the best fit algorithm (with possibly unequal bin sizes).

### 6.1 All Optimal Reduced-Order Approximations are Aggregations

**Theorem 8** *Suppose $\boldsymbol{\phi} \in \mathbb{S}_n, \boldsymbol{\psi} \in \mathbb{S}_m, m < n$, and that $\boldsymbol{\psi}$ is not an aggregation of $\boldsymbol{\phi}$. Then there exists a $\boldsymbol{\psi}' \in \mathbb{S}_m$ such that $d(\boldsymbol{\phi}, \boldsymbol{\psi}') < d(\boldsymbol{\phi}, \boldsymbol{\psi})$.*

The proof of the theorem makes use of a couple of preliminary lemmas.

**Lemma 1** *Suppose $\boldsymbol{\mu} \in \mathbb{R}_+^m, \boldsymbol{\mu} \neq \mathbf{0}$. Then*

$$\sum_{j=1}^m h(\mu_j) = cH((1/c)\boldsymbol{\mu})) + h(c), \tag{26}$$

*where $c = \boldsymbol{\mu}\mathbf{e}_m$ is a normalizing constant.*

**Proof:** We have that

$$\sum_{j=1}^m h(\mu_j) = \sum_{j=1}^m \mu_j \log \frac{1}{\mu_j} = \sum_{j=1}^m \mu_j \log \frac{c}{\mu_j} - \left(\sum_{j=1}^m \mu_j\right) \log c$$

$$= c \sum_{j=1}^m \frac{\mu_j}{c} \log \frac{c}{\mu_j} - c \log c = cH((1/c)\boldsymbol{\mu})) + h(c).$$

This completes the proof. ∎

**Lemma 2** *Suppose $c_1, c_2, b > 0$ with $c_1 + c_2 + b = 1$. For each $\lambda \in [0,1]$, define $\boldsymbol{\psi}(\lambda) \in \mathbb{S}_2$ by*

$$\boldsymbol{\psi}(\lambda) = [\ c_1 + \lambda b \quad c_2 + (1-\lambda)b\ ],$$

*and $G : [0,1] \to \mathbb{R}$ by*

$$G(\lambda) = bH([\lambda\ 1-\lambda]) - H(\boldsymbol{\psi}(\lambda)).$$

*Then*

$$G(\lambda) > \min\{G(0), G(1)\} = \min\{-H(\boldsymbol{\psi}(0)), -H(\boldsymbol{\psi}(1))\}. \tag{27}$$

**Proof:** This follows from elementary calculus. Using the fact that $h'(r) = -(1 + \log r)$ for all $r > 0$, it can be shown that $G(\lambda)$ achieves its maximum at $\lambda^* = c_1/(c_1 + c_2)$ and is decreasing on either side of it. So if $\lambda < \lambda^*$, then $G(\lambda) > G(0)$, whereas if $\lambda > \lambda^*$, then $G(\lambda) > G(1)$. In either case, (27) is satisfied. ∎

**Proof (of Theorem 8):** Suppose $\boldsymbol{\phi} \in \mathbb{S}_n, \boldsymbol{\psi} \in \mathbb{S}_m, m < n$, and $\boldsymbol{\psi}$ is *not* an aggregation of $\boldsymbol{\phi}$. Choose $P \in \mathbb{S}_{n \times m}$ such that $\boldsymbol{\phi}P = \boldsymbol{\psi}$ and $J_{\boldsymbol{\phi}}(P) = V(\boldsymbol{\phi}, \boldsymbol{\psi})$. Since $\boldsymbol{\psi}$ is not an aggregation of $\boldsymbol{\phi}$, at least one row of $P$ contains at least two nonzero (i.e., positive) elements. Let $k$ be such a row, and without loss of generality permute the components of $\boldsymbol{\psi}$ in such a way that $p_{k1} > 0, p_{k2} > 0$. To show that $\boldsymbol{\psi}$ cannot be an optimal approximation of $\boldsymbol{\phi}$ in the $d$ metric, we will construct another distribution $\boldsymbol{\psi}' \in \mathbb{S}_m$ that matches $\boldsymbol{\psi}$ from component 3 onwards. We will do this by perturbing *only* the two elements $p_{k1}, p_{k2}$ in such a way that $p'_{k1} + p'_{k2} = p_{k1} + p_{k2}$, and defining $\boldsymbol{\psi}' = \boldsymbol{\phi}P'$. This means that many of the quantities are common to $\boldsymbol{\psi}$ and $\boldsymbol{\psi}'$, so in the various equations below, we will just write 'const' to avoid notational clutter.

From the manner in which $P$ was chosen, it follows that

$$
\begin{aligned}
V(\boldsymbol{\phi}, \boldsymbol{\psi}) &= \sum_{i=1}^{n} \phi_i H(\mathbf{p}_i) = \phi_k H(\mathbf{p}_k) + \text{const} \\
&= \phi_k [h(p_{k1}) + h(p_{k2})] + \text{const},
\end{aligned}
$$

$$
\begin{aligned}
V(\boldsymbol{\psi}, \boldsymbol{\phi}) &= V(\boldsymbol{\phi}, \boldsymbol{\psi}) + H(\boldsymbol{\phi}) - H(\boldsymbol{\psi}) \\
&= \phi_k [h(p_{k1}) + h(p_{k2})] - [h(\psi_1) + h(\psi_2)] + \text{const}.
\end{aligned}
$$

Note that the 'constant' in the two equations need not be the same. Our use of the phrase 'constant' means only that all the ignored summations remain unchanged when we replace $\boldsymbol{\psi}$ by $\boldsymbol{\psi}'$. Proceeding further, let us write

$$
\psi_1 = \sum_{i=1}^{n} \phi_i p_{i1} = \sum_{i \neq k} \phi_i p_{i1} + \phi_k p_{k1} =: d_1 + \phi_k p_{k1}.
$$

Similarly,

$$
\psi_2 = \sum_{i=1}^{n} \phi_i p_{i2} = \sum_{i \neq k} \phi_i p_{i2} + \phi_k p_{k2} =: d_2 + \phi_k p_{k2}.
$$

With these definitions, we can write

$$
\begin{aligned}
V(\boldsymbol{\psi}, \boldsymbol{\phi}) &= \phi_k [h(p_{k1}) + h(p_{k2})] \\
&\quad - [h(d_1 + \phi_k p_{k1}) + h(d_2 + \phi_k p_{k2})] + \text{const}. \qquad (28)
\end{aligned}
$$

This looks similar to the function $G(\lambda)$ in Lemma 2, except that neither $p_{k1} + p_{k2}$ nor $d_1 + d_2 + \phi_k$ necessarily add up to one. So we proceed as in the proof of Lemma 2 and apply the correction terms from Lemma 1 wherever necessary. Let us define

$$
\lambda = \frac{p_{k1}}{p_{k1} + p_{k2}}, \quad 1 - \lambda = \frac{p_{k2}}{p_{k1} + p_{k2}},
$$

$$
\beta = p_{k1} + p_{k2}, \quad \alpha = d_1 + d_2 + \beta \phi_k,
$$

and note that

$$
\psi_1 + \psi_2 = d_1 + d_2 + \beta \phi_k = \alpha.
$$

With these definitions, and making repeated use of Lemma 1, we get

$$
\phi_k [h(p_{k1}) + h(p_{k2})] = \beta \phi_k H([\lambda \ \ 1 - \lambda]) + \phi_k h(\beta),
$$

$$h(\psi_1) + h(\psi_2) = \alpha H(\boldsymbol{\gamma}) + h(\alpha),$$

where

$$\boldsymbol{\gamma} = \left[\begin{array}{cc} \dfrac{d_1}{\alpha} + \lambda \dfrac{\beta\phi_k}{\alpha} & \dfrac{d_2}{\alpha} + (1-\lambda)\dfrac{\beta\phi_k}{\alpha} \end{array}\right] \in \mathbb{S}_2.$$

Hence

$$V(\boldsymbol{\psi}, \boldsymbol{\phi}) = \alpha\left[\dfrac{\beta\phi_k}{\alpha} H([\lambda \ \ 1-\lambda]) + H(\boldsymbol{\gamma})\right] + \phi_k h(\beta) - h(\alpha) + \text{const}.$$

Now the quantity inside the brackets is like $G(\lambda)$ in Lemma 2, with

$$c_1 = \dfrac{d_1}{\alpha}, c_2 = \dfrac{d_2}{\alpha}, b = \dfrac{\beta\phi_k}{\alpha}.$$

And these three numbers *do* add up to one. So we know from Lemma 2 that the quantity inside the brackets can be made smaller by choosing $\lambda = 0$ or 1. The choice $\lambda = 0$ causes $p_{k1}, p_{k2}, \psi_1, \psi_2$ to be replaced by

$$[p'_{k1} \ p'_{k2}] = [0 \ p_{k1} + p_{k2}], [\psi'_1 \ \psi'_2] = [d_1 \ d_2 + \phi_k(p_{k1} + p_{k2})],$$

while the choice $\lambda = 1$ causes $p_{k1}, p_{k2}, \psi_1, \psi_2$ to be replaced by

$$[p'_{k1} \ p'_{k2}] = [p_{k1} + p_{k2} \ 0], [\psi'_1 \ \psi'_2] = [d_1 + \phi_k(p_{k1} + p_{k2}) \ d_2].$$

In either case the numbers $\beta, \alpha$ remain the same, whence the correction term $\phi_k h(\beta) - h(\alpha)$ also remains the same. So decreasing the quantity inside the brackets reduces $V(\boldsymbol{\psi}, \boldsymbol{\phi})$. So the conclusion is that there exists a $P' \in \mathbb{S}_{n \times m}$ such that, with $\boldsymbol{\psi}' = \boldsymbol{\phi} P'$, we have

$$\begin{aligned} J_{\boldsymbol{\phi}}(P') + H(\boldsymbol{\phi}) - H(\boldsymbol{\psi}') &= \phi_k[h(p'_{k1}) + h(p'_{k2})] - [h(\psi'_1) + h(\psi'_2)] + \text{const} \\ &< \phi_k[h(p_{k1}) + h(p_{k2})] - [h(\psi_1) + h(\psi_2)] + \text{const} \\ &= V(\boldsymbol{\psi}, \boldsymbol{\phi}). \end{aligned}$$

Now, since $V(\boldsymbol{\phi}, \boldsymbol{\psi}')$ is the *minimum* of the quantity $J_{\boldsymbol{\phi}}(Q)$ over all $Q \in \mathbb{S}_{n \times m}$ such that $\boldsymbol{\phi} Q = \boldsymbol{\psi}'$, we conclude from the above that

$$V(\boldsymbol{\phi}, \boldsymbol{\psi}') + H(\boldsymbol{\phi}) - H(\boldsymbol{\psi}') < V(\boldsymbol{\psi}, \boldsymbol{\phi}),$$

or equivalently that

$$V(\boldsymbol{\psi}', \boldsymbol{\phi}) < V(\boldsymbol{\psi}, \boldsymbol{\phi}).$$

Similarly, we can compare $V(\boldsymbol{\phi}, \boldsymbol{\psi}')$ and $V(\boldsymbol{\phi}, \boldsymbol{\psi})$. Since $p'_{k1} + p'_{k2} = p_{k1} + p_{k2}$, and one of $p'_{k1}, p'_{k2}$ is zero, it is obvious that

$$\phi_k[h(p'_{k1}) + h(p'_{k2})] < \phi_k[h(p_{k1}) + h(p_{k2})].$$

Hence
$$J_\phi(P') < J_\phi(P) = V(\phi, \psi).$$

As a consequence, we have as before that
$$V(\phi, \psi') = \min_{Q \in \mathbb{S}_{n \times m}} J_\phi(Q) \text{ s.t. } \phi Q = \psi' \leq J_\phi(P') < V(\phi, \psi).$$

Combining both inequalities leads to $d(\phi, \psi') < d(\phi, \psi)$. ∎

## 6.2 Greedy Algorithm for Finding Suboptimal Solution

Theorem 8 shows that the best reduced-order approximations in $\mathbb{S}_m$ to the given $\phi \in \mathbb{S}_n$ is an aggregation of $\phi$. Now, suppose $\phi^{(a)} \in \mathbb{S}_m$ is an aggregation of $\phi$. Then it is clear that $V(\phi, \phi^{(a)}) = 0$. Hence it follows from (7) that
$$d(\phi, \phi^{(a)}) = V(\phi^{(a)}, \phi) = H(\phi) - H(\phi^{(a)}).$$

Hence an aggregation $\phi^{(a)}$ of $\phi$ into $m$ states is an optimal approximation if and only if $\phi^{(a)}$ has maximum entropy amongst all aggregations. Note that when $m = 2$, an aggregation $\phi^{(a)}$ has maximum entropy if and only if the total variation $\rho(\phi^{(a)}, \mathbf{u}_2)$ is minimized, where $\mathbf{u}_m$ denotes the uniform distribution with $m$ components. This is a bin-packing problem with overstuffing where both bins have capacity 0.5 and is thus NP-hard. It is plausible that the problem remains NP-hard for $m \geq 3$ as well. So we reformulate the problem. Amongst all distributions in $\mathbb{S}_m$, the uniform distribution $\mathbf{u}_m$ has the maximum entropy. Thus we seek an aggregation of $\phi$ in such a way that the total variation distance $\rho(\phi^{(a)}, \mathbf{u}_m)$ is minimized. This is yet another bin-packing problem with overstuffing, with all bin sizes equal to $1/m$. A suboptimal aggregation can therefore be found using the best fit algorithm, whose complexity is $O(n \log m)$ as discussed earlier.

Therefore when $m = 2$ the optimal-reduced order approximation problem is NP-hard, so we are justified in seeking an efficient suboptimal algorithm even when $m \geq 2$.

We complete this section with an upper bound on the performance of the best fit algorithm, *without* assuming that the bin sizes are equal. This result may be of independent interest. By specializing to the case where $\psi = \mathbf{u}_m$, this bound can be translated into a bound on the entropy of the resulting aggregation. The details are easy and left to the reader.

**Theorem 9** *For the best fit algorithm with bin size vector $\psi$, we have*
$$\rho(\phi^{(a)}, \psi) \leq 0.25 m \phi_{\max}. \tag{29}$$

*where $\phi_{\max} = \max_i \{\phi_i\}$.*

**Proof:** The steps in the proof follow the corresponding steps in [18]. Once the greedy algorithm is completed, let us denote the resulting aggregation $\boldsymbol{\phi}^{(a)}$ by $\boldsymbol{\alpha}$ to reduce clutter. Let us refer to bin $j$ as 'heavy' if $\alpha_j > \psi_j$, and 'light' if $\alpha_j \leq \psi_j$. Suppose there are $k$ heavy bins. Without loss of generality, renumber the bins such that the first $k$ bins are heavy and the rest are light. Let $e_1, \ldots, e_k$ denote the excess and $s_{k+1}, \ldots, s_m$ denote the slack. In other words,

$$e_j = \alpha_j - \psi_j, j = 1, \ldots, k, \text{ and } s_j = \psi_j - \alpha_j, j = k+1, \ldots, m.$$

For $j = 1, \ldots, k$, let $r_j$ denote its excess capacity *just before* the last item was placed into it (making it heavy). Then two things are obvious. First, $r_j + e_j$ equals the last component of $\boldsymbol{\phi}$ that was placed into this bin, and as a result $r_j + e_j \leq \phi_{\max}$. Second, the nature of the algorithm implies that $r_j$ is at least equal to the capacity of all the other bins at the time this item was placed into bin $j$. Since bin capacity can only decrease as the algorithm is run, in particular this implies that

$$r_j \geq s_{k+1}, \ldots, s_m, j = 1, \ldots, k.$$

Therefore

$$\frac{1}{k} \sum_{j=1}^{k} r_j \geq \min_{j=1,\ldots,k} r_j \geq \max_{j=k+1,\ldots,m} s_j \geq \frac{1}{m-k} \sum_{j=k+1}^{m} s_j.$$

Rearranging gives

$$(m-k) \sum_{j=1}^{k} r_j \geq k \sum_{j=k+1}^{m} s_j.$$

Since both $\boldsymbol{\phi}, \boldsymbol{\psi}$ are unit vectors, it follows that

$$\sum_{j=1}^{k} e_j = \sum_{j=k+1}^{m} s_j.$$

Therefore

$$(m-k) \sum_{j=1}^{k} (r_j + e_j) \geq m \sum_{j=k+1}^{m} s_j.$$

Note that the right side is precisely $m\rho(\boldsymbol{\alpha}, \boldsymbol{\psi})$. Hence

$$\rho(\boldsymbol{\alpha}, \boldsymbol{\psi}) \leq \frac{m-k}{m} \sum_{j=1}^{k} (r_j + e_j) \leq \frac{k(m-k)}{m} \phi_{\max} \leq \frac{m\phi_{\max}}{4},$$

which follows from the obvious observation that $k(m-k) \leq m^2/4$ no matter what $k$ is.

It is quite easy to show that the performance of the algorithm is bounded by $0.5m\phi_{\max}$. This is because no bin can be overstuffed by more than $\phi_{\max}$, and no bin can have unutilized capacity more than $\phi_{\max}$. Since the totals of over- and under-capacity have to balance out, the bound $0.5m\phi_{\max}$ follows. Thus the real essence of the theorem is to gain an extra factor of 0.5.

The specific result of [18] bounds the *total weight* of all the bins (call it $A$) and shows that $A \leq 1.25 A_{\text{opt}}$. Moreover, the proof also requires an extra assumption that $\phi_{\max} \leq \psi_{\min}$, something that is not needed here. It is easy to verify that the weight of an algorithm equals $1 + \rho(\phi^{(a)}, \psi)$ achieved by that algorithm. Hence a direct application of the results of [18] would imply that

$$\rho(\phi^{(a)}, \psi) \leq 1.25 \rho_{\text{opt}}(\phi^{(a)}, \psi) + 0.25.$$

Because of the additive constant of 0.25, this bound is less useful than the bound (29) given by Theorem 9.

## 6.3 Example of Aggregation Using the Greedy Algorithm

**Example 3** *Let us return to the 40-dimensional probability distribution $\phi$ studied in Example 2. As seen earlier, $H(\phi) = 3.6399$, while the maximum entropy that any 10-dimensional distribution can have is $\log(10) \approx 2.3026$.*

*Applying the best fit algorithm for aggregation without sorting $\phi$ results in the following grouping and aggregation (shown as a matrix for convenience):*

$$I_1 = \{1, 16, 23, 33\}, I_2 = \{2, 18, 31\}, I_3 = \{3, 12, 24, 40\}, I_4 = \{4, 19, 30, 36\},$$

$$I_5 = \{5, 15, 27, 38\}, I_6\{6, 11, 17, 25, 34\}, I_7 = \{7, 13, 26, 37\}, I_8 = \{8, 14, 22, 28, 35\},$$

$$I_9 = \{9, 20, 32, 39\}, I_{10} = \{10, 21, 29\},$$

$$\phi^{(a)} = \begin{bmatrix} 0.0951 & 0.0942 & 0.0989 & 0.1099 & 0.1020 \\ 0.0917 & 0.1085 & 0.0938 & 0.1188 & 0.0871 \end{bmatrix}.$$

*We have that $H(\phi^{(a)}) = 2.2934$, quite close to the theoretical maximum of 2.3026.*

*In contrast, if we first sort $\phi$ before applying the best fit algorithm, the following grouping results:*

$$I_1 = \{1, 20, 21, 39\}, I_2 = \{2, 19, 22, 40\}, I_3 = \{3, 18, 23, 38\}, I_4 = \{4, 17, 24, 37\},$$

$$I_5 = \{5, 16, 25, 36\}, I_6 = \{6, 15, 29, 31\}, I_7 = \{7, 14, 27, 34\}, I_8 = \{8, 13, 26, 35\},$$

$$I_9 = \{9, 12, 28, 33\}, I_{10} = \{10, 11, 30, 32\}.$$

*The resulting aggregation is*

$$\boldsymbol{\phi}^{(a)} = \begin{bmatrix} 0.1019 & 0.1021 & 0.1016 & 0.1007 & 0.1004 \\ 0.0982 & 0.0994 & 0.0993 & 0.0982 & 0.0982 \end{bmatrix},$$

*which is much closer to being uniform than the earlier aggregation.*

# 7 Conclusions

In this paper we have studied the problem of defining a metric distance between two probability distributions over distinct finite sets of possibly different cardinalities. Along the way, we have formulated the problem of constructing a joint distribution on the product of the two sets, which has the two given distributions as its marginals, in such a way that the joint distribution has minimum entropy. While the problem of maximizing mutual information is occasionally discussed in the literature, this specific problem does not appear to have been studied earlier. This problem turns out to be NP-hard, so we reformulated the problem as a bin-packing problem with overstuffing, and adapt the best fit algorithm for bin-packing, leading to an upper bound on the distance between the two given distributions. The complexity of this algorithm is $O((n + m^2) \log m)$, where $n$ is the larger of the two cardinalities and $m$ is the smaller.

Once the metric is defined, we then study the problem of optimal order reduction, namely, given an $n$-dimensional distribution, finding the (or a) $m$-dimensional distribution that is closest in the metric to the given distribution. This turns out to be equivalent to aggregating the original distribution so as to maximize the entropy of the aggregated distribution. This problem is also NP-hard. Accordingly, the problem is again formulated as a bin-packing problem with overstuffing. A greedy algorithm with complexity $O(n \log m)$ is presented, and an upper bound is derived for the error of the greedy algorithm.

The work described here is in contrast with earlier work [8, 9, 10, 17] in which the attempt is to define a notion of *divergence*, not necessarily a metric, between two distributions on distinct sets.

There are several important follow-up problems that arise from the work reported here. The distance between two probability distributions $\boldsymbol{\phi}, \boldsymbol{\psi}$ can be thought of as the distance rate between two i.i.d. processes whose one-dimensional marginals are these two distributions. So the results presented here permit the approximation of an i.i.d. process over a set of large cardinality by another i.i.d. process over a set of smaller cardinality. In order to be truly useful in the context of control over networks for example, the next logical step is to extend the theory to finite-state Markov chains, and finite-state hidden Markov models. We propose to tackle this extension in future research. If we succeed in that, then the next logical step would be to extend the work still further to arbitrary stationary ergodic processes, using the Shannon-McMillan-Breiman theorem.

# Acknowledgements

The author thanks the reviewers of the first version of the paper for several useful comments that helped to improve the presentation. He also thanks Soura Dasgupta and Yutaka Yamamoto for helpful suggestions on this version.